\documentclass[12pt]{iopart}
\usepackage{graphics,graphicx,subfigure,adjustbox,xcolor}

\begin{document}

\title[Disordered ground states in 1D exactly solvable fluids]{Disordered ground states in one-dimensional exactly solvable fluids}

\author{Igor Trav\v{e}nec and Ladislav \v{S}amaj} 

\address{Institute of Physics, Slovak Academy of Sciences, 
D\'ubravsk\'a cesta 9, SK-84511 Bratislava, Slovakia}
\ead{Igor.Travenec@savba.sk,Ladislav.Samaj@savba.sk}
\vspace{10pt}
\begin{indented}
\item[]
\end{indented}

\begin{abstract}
One-dimensional fluids of classical hard-core particles of diameter $a$,
interacting in pairs via a soft repulsive (monotonically decreasing)
potential of finite range 
$\varphi(x)=\varepsilon \left[ (a'-x)/(a'-a)\right]^{1/\nu}$ $(a\le x\le a')$
with real positive parameters $\varepsilon$ and $\nu$, are studied in
an isothermal-isobaric ensemble.
If $a'\le 2a$, the pairwise interactions are reduced to nearest-neighbour
interactions, which allows for an exact solution of the thermal equilibrium.
We focus on the $T\to 0$ ground states, specifically on the equation of state
for the mean distance between nearest neighbours $l_0$ and the pair correlation
function $g_0(x)$.
If $\varphi(x)$ is concave $(\nu\ge 1)$, there exists an ``incompressibility''
pressure $p_i=\varepsilon/(a'-a)$ such that the ground state is
an equidistant chain of particles with spacing $l_0=a'$ for $0<p<p_i$
and with spacing $l_0=a$ for $p>p_i$.
If $\nu>1$ (strict concavity), the ground state at $p=p_i$ is disordered
with $l_0=\left[ \nu a +(\nu-1)a'\right]/(2\nu-1)$ and $g_0(x)$ being
a superposition of weighted Dirac delta functions over discrete positions.
If $\nu=1$ (linear ramp), the ground state at $p=p_i$ is disordered
with $l_0=(a+a')/2$ and the continuous $g_0(x)$ is a superposition of
Heaviside step functions multiplied by polynomials in $x$.
The isothermal susceptibility at $T=0$ is nonzero for concave
$\varphi(x)$ ($\nu\ge 1$) at $p=p_i$ and, therefore, the corresponding
disordered ground states are non-hyperuniform, i.e., they
resemble disordered fluids at nonzero temperatures.
It turns out that pair correlation functions of disordered
ground states, which occur only at a single pressure $p_i$, extend their
predictive power to thermodynamic states at nonzero temperatures over
a wider range of pressures around $p_i$.
If $\nu\in (0,1)$ (strict convexity of the soft potential),
there are no disordered states.

\end{abstract}


\vspace{2pc}


\submitto{\JPA}

\maketitle

\renewcommand{\theequation}{1.\arabic{equation}}
\setcounter{equation}{0}

\section{Introduction} \label{Sec1}
According to the ``crystallization conjecture''
\cite{Radin87,Bris05,Blanc15,Betermin21}, the ground states of classical
many-particle systems are usually crystalline, in agreement with
Nernst's third law of thermodynamics. 
However, there are equilibrium statistical systems with specific topology of
pairwise interactions that remain disordered down to zero temperature.
Such systems generally have a large number of ground states
with the same energy and lack periodic ordering.

A typical example of disorder systems at zero temperature are interacting
classical spins on a lattice, see recent reviews \cite{Capponi25,Chen25}.
The exact solution of the antiferromagnetic Ising model on a frustrated
triangular lattice \cite{Wannier50,Houtappel50,Wannier73} shows
that the ground state is disordered and has finite entropy,
whereas the long-range pair correlations exhibit oscillations and decay as
an inverse power law for large distances \cite{Stephenson64,Stephenson70}.
The number of fully packed dimer configurations was determined on planar
graphs in Refs. \cite{Kasteleyn61,Temperley61}. 
Lieb's exact solution of a vertex model on the square lattice with
the ice condition of having precisely two arrows pointing into each vertex
also gives a nonzero residual entropy at zero temperature \cite{Lieb67}.
Disordered ground states can arise also for one-dimensional (1D)
chains of many-state spins interacting with a group of spins whose
Hamiltonians are constrained by either spin or spatial inversion symmetry
\cite{Radin83,Canright96}.

There are fluid systems (formulated on a continuous space)
of classical particles interacting via a pair potential that are disordered
at zero temperature.
Such systems can be of two kinds \cite{Torquato18}.
Density fluctuations are suppressed on large length scale in {\em hyperuniform}
systems. 
Mathematically, the Fourier component of the structure factor $\tilde{S}(k)$
approaches zero as $k\to 0$. 
According to this property, hyperuniform ground states structurally resemble
crystals and quasicrystals, for which also $\lim_{k\to 0}\tilde{S}(k)=0$.
For {\em non-hyperuniform} ground states, the structure factor
$\tilde{S}(k)>0$ in the limit $k\to 0$ and the systems resemble disordered
liquids at nonzero temperature. 

Hyperuniform ground states have attracted interest in recent years.
Their structure factor $\tilde{S}(k)$ vanishes in the limit $k\to 0$ either
as $k^{\alpha}$ $(\alpha>0)$ or $\tilde{S}(k)=0$ for $0\le k \le K$ with $K>0$
(incident radiation is entirely unscattered in this interval of $k$);
in the latter case hyperuniform systems are referred to as ``stealthy''.
Classical systems of particles interacting via a soft
absolutely integrable potential whose Fourier component is positive,
bounded, and has compact support at some finite wavenumber have
been shown to exhibit stealthy hyperuniform ground states above a critical
density \cite{Fan91,Uche04,Uche06,Batten08,Batten09a,Batten09b}.
In Euclidean space, these soft potentials are isotropic, bounded,
oscillatory, and exhibit inverse power decay at large distances.
An ensemble theory for stealthy disordered ground states has been constructed
\cite{Torquato15}.
Hyperuniform systems of particles with a unique disordered ground state
have been proposed in Refs. \cite{Zhang16,Zhang17}.

In this paper we consider 1D fluids of classical particles interacting
by repulsive potentials of finite range.
In the case of {\em convex} repulsive potentials, Ventev\"ogel proved
the existence of a periodic ground state for any particle density
\cite{Ventevogel78}.
Similar theorems have been derived for certain {\em non-convex} repulsive
potentials \cite{Ventevogel79a,Ventevogel79b}.

There exists a family of 1D classical fluids of particles with pairwise
interactions which effectively reduce themselves to the nearest-neighbour
(NN) interactions.
The thermal equilibrium of such systems, namely thermodynamic potentials,
the Equation of State (EoS), the structure factor, and the correlation
functions, is exactly solvable.
The first and simplest model of this kind was the Tonks gas of hard rods
defined by the interaction potential
\begin{equation}\label{hr}
\phi(x) = \left\{
\begin{array}{ll}
\infty & \mbox{if $\vert x\vert < a$,} \cr
0 & \mbox{if $\vert x\vert \ge a$,}  
\end{array} \right.   
\end{equation}
where $a$ is the diameter of the hard core.
Its EoS was derived by Tonks \cite{Tonks36}, two-particle distribution
function was calculated sooner by Zernike and Prins \cite{Zernike27}.
Takahashi \cite{Takahashi42} proposed a more general model
\begin{equation} \label{general}
\phi(x) = \left\{
\begin{array}{ll}
\infty & \mbox{if $\vert x\vert < a$,} \cr
\varphi(\vert x\vert) & \mbox{if $a\le \vert x\vert < a'$,} \cr
0 & \mbox{if $\vert x\vert \ge a'$,}  
\end{array} \right.   
\end{equation}
with two competing length scales: the hard-core diameter $a$ and
the finite range $a'>a$ of the soft component $\varphi(x)$
of the interaction potential.
Assuming ad-hoc that particles interact only with their NNs,
Takahashi derived the EoS using canonical ensemble.
A rederivation of the EoS using the isothermal-isobaric ensemble
was performed by Bishop and Boonstra \cite{Bishop83}.
The many-particle distribution functions were
computed by Salsburg, Zwanzig and Kirkwood \cite{Salsburg53}.
An alternative formula for the two-particle distribution function was derived
in \cite{Lebowitz62}.
The exact solution includes mixtures of particles, e.g. nonadditive
hard rods \cite{Lebowitz71,Heying04,Santos07,Ben-Naim09,Sahnoun24},
see the review \cite{Percus87} and chapter 5 of the monograph \cite{Santos16}.
Takahashi's original idea gained a rigorous basis by introducing a condition
for the finite range of the soft potential $a'$ 
in (\ref{general}):
\begin{equation} \label{condition}
a'\le 2a .
\end{equation}
This constraint limits the interaction of each particle only to its NNs,
without any ad-hoc assumption; the interaction with next-to-NN particles
is equal to zero because of the limiting hard cores of the NN particles.
The condition (\ref{condition}) will be automatically taken into account
throughout this study.

In a recent work \cite{Travenec25} we studied 1D classical fluids with
repulsive NN interactions in the context of negative thermal expansion
(NTE), that is, volume shrinkage upon isobaric heating of the system.
It was shown that this low-temperature anomaly occurs in many cases
of the interaction potential for which the EoS can be explicitly obtained
in terms of elementary or special functions like the square shoulder,
the linear ramp, the fusion of linear and quadratic ramps and
certain logarithmic potentials.
In two specific models, namely the square shoulder and the linear ramp,
NTE was related to a jump in particle density $n$ between two
equidistant ground states with periods $a'$ and $a$ at
an ``incompressibility'' pressure $p_i$ (beyond which the system is no
longer compressible): $n=1/a'$ for $0<p<p_i$ and $n=1/a$ for $p>p_i$.
For $p=p_i$ we got $n=2/(a+a')$ for both models.
This result suggested that the ground state at $p=p_i$ is a simple mixture
of particle chains, with one half having NN distances $a$ and the other half
having NN distances $a'$.
Therefore, the nature and characteristics of these ground states were not
analysed in Ref. \cite{Travenec25}.

In this paper, we study a broader family of concave and
convex soft potentials, which includes the square shoulder and linear ramp NN
models as special (concave) cases.
It is shown that concave potentials exhibit rather complex disordered states
at $p=p_i$.
These states deserve attention for their exact solvability, which is
an exceptional phenomenon in the field of disordered systems.
Obtaining these disordered states from the exact EoS and equations for the
correlation functions is nontrivial and requires mathematical manipulations
that go beyond those presented in Ref. \cite{Travenec25}.
For the square shoulder potential, the correlation function at incompressibility
pressure $p_i$ turns out to be a superposition of weighted Dirac delta
function, while for the linear ramp potential it is a superposition of
Heaviside step functions multiplied by polynomials in distance.
It turns out that the pair correlation functions of disordered states
extend their predictive power to thermodynamic states at nonzero temperatures
over a wider range of pressures around $p_i$.
The isothermal susceptibility is nonzero in disordered ground states
or, equivalently, the structure function $\tilde{S}(k)>0$ in the limit $k\to 0$.
The disordered ground states are therefore non-hyperuniform, i.e., they
look like standard liquids at nonzero temperatures.
For strictly convex interaction potentials, there are no disordered ground
states.

\begin{figure}[t]
\begin{center}
\includegraphics[clip,width=0.84\textwidth]{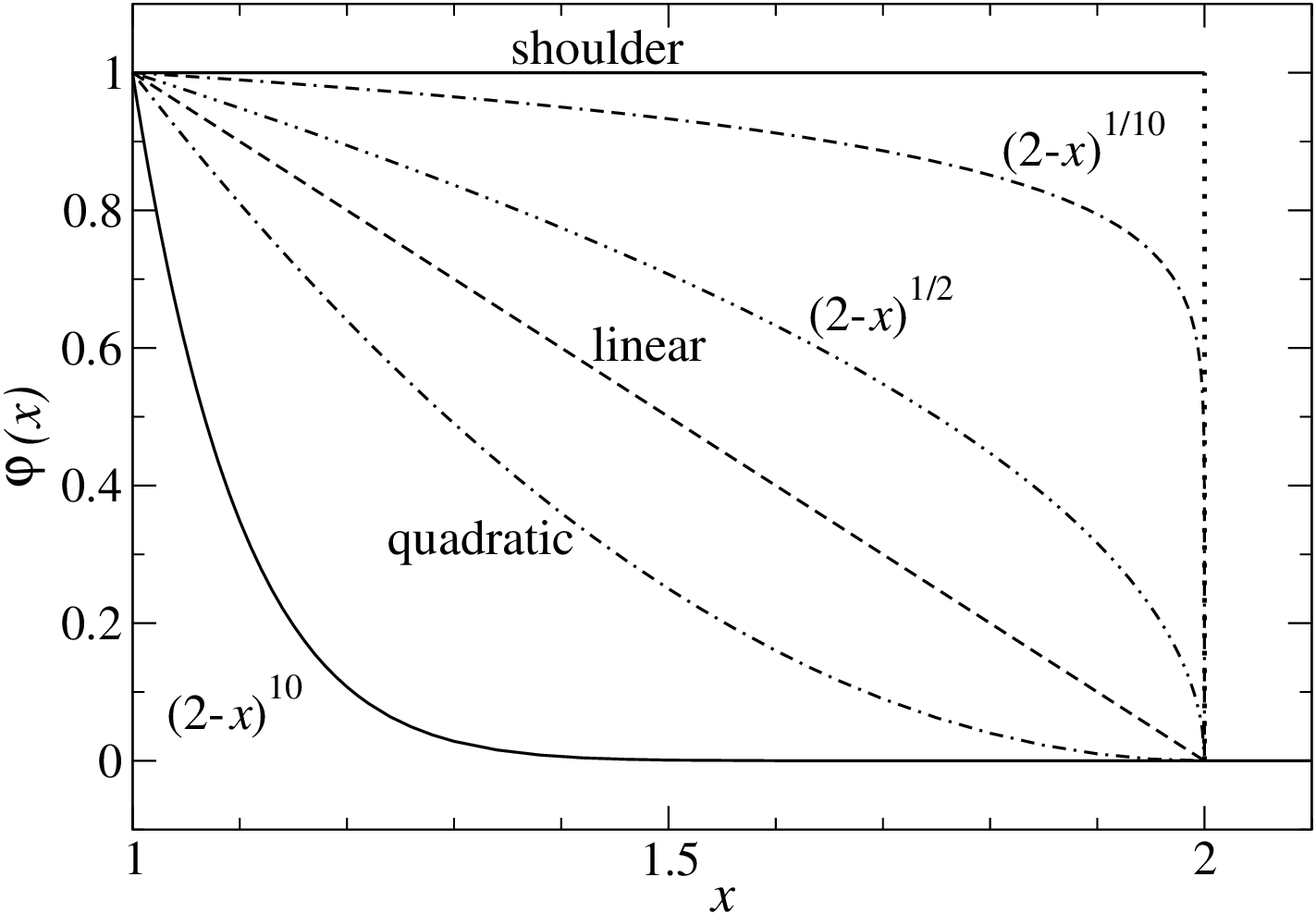}
\caption{Some of the interaction potentials (\ref{nu}), with parameters $a=1$,
$a'=2$ and $\varepsilon=1$.
The case $\varphi=1$ is denoted as shoulder, $\varphi=2-x$ as linear
and $\varphi=(2-x)^2$ as quadratic.}
\label{fig1}
\end{center}
\end{figure}

The set of repulsive soft potentials in (\ref{general}) with the condition
(\ref{condition}) that we consider in this paper is
\begin{equation} \label{nu}
\varphi(x) = \varepsilon \left( \frac{a'-x}{a'-a} \right)^{1/\nu} ,
\qquad a\le x \le a',
\end{equation}
where the real parameters $\varepsilon$ and $\nu$ are positive.
These interaction potentials are monotonically decreasing functions of
distance $x$, starting from $\varepsilon>0$ at the hard-core point $x=a$
and vanishing at the point $x=a'$, then continuing with the value $0$
for $x>a'$. 
Some of the potentials with parameters fixed at $a=1$, $a'=2$ and
$\varepsilon=1$ are shown in figure \ref{fig1}.
The potentials are strictly concave if $\nu>1$, the limit $\nu\to\infty$
corresponds to a square shoulder.
The potentials are strictly convex if $0<\nu<1$.
The case $\nu=1$ corresponds to a linear ramp that is both concave and
convex.

The paper is organized as follows.
Section \ref{Sec2} deals with a general formalism for exactly solvable
1D fluids with NN interactions, at an arbitrary temperature $T$ and in
the ground state at $T=0$.
The explicit formulas for the EoS, pair correlation function and
isothermal susceptibility are presented.
The results for the simplest Tonks model (\ref{hr}) are briefly summarized.
Strictly concave $(\nu>1)$ ramp potentials are discussed
in section \ref{Sec3}.
Section \ref{Sec4} deals with the linear $\nu=1$ ramp.
Strictly convex $(0<\nu<1)$ ramp potentials are analysed
in section \ref{Sec5}. 
The final section \ref{Sec6} recapitulates the study.

\renewcommand{\theequation}{2.\arabic{equation}}
\setcounter{equation}{0}

\section{General formalism} \label{Sec2}

\subsection{Exact formulas for NN fluid} \label{Sec2.1}
$N$ identical particles move in a box of length $L$, say with periodic
boundary conditions.
The thermodynamic limits $L\to\infty$ and $N\to\infty$ are considered,
while the particle number density $n=N/L$ remains fixed. 
The pair potential $\phi(x)$ is of type (\ref{general}) with the NN condition
$a'\le 2 a$.
The condition $\lim_{x\to 0}\phi(x)=\infty$ fixes the particle order
on the line.
The particles are in thermal equilibrium at temperature $T$,
or the inverse temperature $\beta=1/(k_{\rm B}T)$, where
$k_{\rm B}$ is the Boltzmann constant.
The model is exactly solvable in the isothermal-isobaric ensemble,
for notation see the monograph \cite{Santos16}.

The particle density at position $x$ is given by
\begin{equation}
n(x) = \langle \sum_j \delta(x-x_j) \rangle ,
\end{equation}
where the sum is over all particles at positions $\{ x_j\}_{j=1}^N$ and
the average $\langle \cdots\rangle$ is taken over the isothermal-isobaric
ensemble with a fixed temperature $T$ and pressure $p$ that
acts on the system. 
The particle density does not depend on $x$, $n(x)=n$.
The exact solution for EoS can be written using the Laplace transform of
the Boltzmann factor ${\rm e}^{-\beta\phi(x)}$,
\begin{equation} \label{Om}
\widehat{\Omega}(s) = \int_0^{\infty} {\rm d}x\,
{\rm e}^{-xs} {\rm e}^{-\beta\phi(x)} 
\end{equation}
and its derivative
\begin{equation}\label{om'}
\widehat{\Omega}'(s) \equiv \frac{\partial\widehat{\Omega}(s)}{\partial s}
= - \int_0^{\infty} {\rm d}x\, x {\rm e}^{-xs} {\rm e}^{-\beta\phi(x)}
\end{equation}  
as follows
\begin{equation} \label{EoS}
n(T,p) = - \frac{\widehat{\Omega}(\beta p)}{\widehat{\Omega}'(\beta p)}
= \frac{\int_0^{\infty} {\rm d}x\, {\rm e}^{-x\beta p} {\rm e}^{-\beta\phi(x)}}{
\int_0^{\infty} {\rm d}x\, x {\rm e}^{-x\beta p} {\rm e}^{-\beta\phi(x)}} . 
\end{equation}  
By introducing the average distance between NN particles
$l(T,p)=1/n(T,p)$ we can alternatively write
\begin{equation} \label{lEoS}
l(T,p) = - \frac{\partial}{\partial s}
\ln \widehat{\Omega}(s+\beta p) \Bigg\vert_{s=0} . 
\end{equation}
The ground state spacing is obtained as the limit $T\to 0$ of $l(T,p)$.
In what follows, we will use the notation $l_0(p)\equiv l(0,p)$.
The ``incompressibility'' pressure $p_i$ is defined by the condition
\begin{equation}  
l_0(p) = a \qquad \mbox{for all $p>p_i$,}
\end{equation}
i.e. the corresponding equidistant ground state consists of the
close-packed array of hard rods.

The pair correlation function $g(x,x')$ between two points is equal to
the ratio
\begin{equation}
g(x,x') = \frac{n^{(2)}(x,x')}{n(x) n(x')} ,
\end{equation}  
where
\begin{equation}
n^{(2)}(x,x') = \langle \sum_{j\ne k} \delta(x-x_j) \delta(x'-x_k) \rangle .
\end{equation}  
The correlation function depends on the distance between the points $x$ and
$x'$, $g(x,x') = g(\vert x-x'\vert)$, and is equal to 0 if this distance is
smaller than or equal to $a$, $g(x)=0$ if $x\le a$.
The Laplace transform of $g(x)$, defined as
\begin{equation}
\widehat{G}(s) \equiv \int_0^{\infty} {\rm d}x\, {\rm e}^{-x s} g(x) ,
\end{equation}  
is given as \cite{Santos16}
\begin{equation} \label{G}
\widehat{G}(s) = \frac{1}{n}\ \frac{\widehat{\Omega}(s+\beta p)}{
\widehat{\Omega}(\beta p)-\widehat{\Omega}(s+\beta p)}
= \frac{1}{n}\sum_{k=1}^\infty
\left[\frac{\widehat{\Omega}(s+\beta p)}{\widehat{\Omega}(\beta p)}\right]^k .
\end{equation}
The normalization condition for $g(x)$ is
\begin{equation} \label{norm}
\lim_{L\to\infty}\frac{1}{L-a}\int_a^L g(x){\rm d}x = 1.
\end{equation}

The static structure factor $\tilde{S}(k)$ is defined as Fourier transform
\begin{eqnarray}
\tilde{S}(k) & = & 1 + n \int_{-\infty}^{\infty} {\rm d}x\, {\rm e}^{-{\rm i} k x}
\left[ g(x)-1 \right] \nonumber \\ & = &  1 + 2 n \int_0^{\infty} {\rm d}x\,
\cos(k x) \left[ g(x)-1 \right] . 
\end{eqnarray}  
The isothermal susceptibility is given by
\begin{equation}\label{chi}
\chi_T = \left( \frac{\partial{n}}{\partial {\beta p}}\right)_\beta
= -1+\frac{\widehat{\Omega}(\beta p)\widehat{\Omega}''(\beta p)}{
\left[\widehat{\Omega'}(\beta p)\right]^2} .
\end{equation}
It is related to the pair correlation function as follows \cite{Santos16}
\begin{equation} \label{chig}
\chi_T = 1 + n \int_{-\infty}^{\infty}{\rm d}x\, \left[ g(x)-1\right]
= \tilde{S}(0) .
\end{equation}
This means that the condition of hyperuniformity $\tilde{S}(0)=0$ is
equivalent to the condition $\chi_T=0$.

\subsection{Tonks gas} \label{Sec2.4}
In this section we consider the exact solution of the simplest 1D fluid of hard
rods with an interaction potential (\ref{hr}).
The Laplace transform of the Boltzmann factor (\ref{Om}) becomes
\begin{equation}
\widehat{\Omega}(s) = \frac{{\rm e}^{-a s}}{s} .
\end{equation}
The EoS for the reciprocal density is
\begin{equation} \label{EoShr}
l(T,p) \equiv \frac{1}{n} = a + \frac{T}{p} ,
\end{equation}
where $l$ is the mean distance between the NN pairs of particles.
In the low-temperature limit $T\to 0$ and for any positive pressure $p>0$
we have $l_0(p)=a$, that is, the incompressibility pressure $p_i=0$.  

Let us reproduce this result by directly calculating the pair
correlation function at zero temperature.
By substituting
\begin{equation} \label{om/om}
\lim_{\beta\to\infty}\frac{\widehat{\Omega}(s+\beta p)}{
\widehat{\Omega}(\beta p)} =\lim_{\beta\to\infty}
\frac{\beta p}{\beta p+s}{\rm e}^{-a s}={\rm e}^{-a s} 
\end{equation}
into (\ref{G}) and performing the inverse Laplace transformation,
the distribution function at zero temperature is
\begin{equation} \label{g0hr}
g_0(x) = {\cal L}^{-1}\left\{a\sum_{k=1}^\infty{\rm e}^{-a k s}\right\}
=a\sum_{k=1}^{\infty} \delta(x-k a),
\end{equation}
where $\delta(x)$ is Dirac delta function.
This is an equidistant ground state with lattice constant $a$.

Regarding the isothermal susceptibility (\ref{chi}), for $T\to 0$ we find
that the susceptibility
\begin{equation}\label{chi0t}
\chi_0=\lim_{\beta\to\infty}\frac{1}{(1+a\beta p)^2}=0 
\end{equation}
vanishes for the equidistant ground state.

\renewcommand{\theequation}{3.\arabic{equation}}
\setcounter{equation}{0}

\section{Strictly concave interaction potentials with $\nu>1$} \label{Sec3}
Let us consider strictly concave interaction potentials (\ref{nu}) with $\nu>1$.
As follows from the formulas for the EoS (\ref{lEoS}) and the Laplace
transform of the correlation function (\ref{G}), the quantity of interest is
\begin{eqnarray}
\widehat{\Omega}(s+\beta p) & = & \int_a^{a'} {\rm d}x\,
{\rm e}^{-xs} {\rm e}^{-\beta[px+\varphi(x)]} +
\int_{a'}^{\infty} {\rm d}x\, {\rm e}^{-x(s+\beta p)} \nonumber \\
& = & I_{\beta}(a,a';p,s) + \frac{1}{s+\beta p} {\rm e}^{-a'(s+\beta p)} ,
\label{rovnica}
\end{eqnarray}
where
\begin{equation} \label{Iaaps}
I_{\beta}(a,a';p,s) \equiv \int_a^{a'} {\rm d}x\, {\rm e}^{-xs} {\rm e}^{-\beta f(x)}
\end{equation}
with
\begin{equation} \label{fx}
f(x) = p x + \varphi(x) = p x +
\varepsilon \left( \frac{a'-x}{a'-a} \right)^{1/\nu} . 
\end{equation}
The dependence of $f(x)$ on $p$ is omitted for simplicity.

The integral $I_{\beta}(a,a';p,s)$ will be treated in the limit $\beta\to\infty$
by the steepest descent method.
This requires determining the point $x_{\min}$ from the interval $[a,a']$
at which $f(x)$ reaches its minimum.
The function $f(x)$ has an extreme value at the point $x^*$ such that
its first derivative $f'(x^*)=0$, i.e.
\begin{equation}
x^* = a' - \left( \frac{\varepsilon}{p\nu} \right)^{\nu/(\nu-1)}
\frac{1}{(a'-a)^{1/(\nu-1)}} .  
\end{equation}
Since
\begin{equation}
f''(x^*) = - \frac{\nu-1}{\nu} (a'-a)^{\frac{1}{\nu-1}} p^{\frac{2\nu-1}{\nu-1}}
\left( \frac{\nu}{\varepsilon} \right)^{\frac{\nu}{\nu-1}}  
\end{equation}
is negative, the extremum at point $x^*$ is the maximum of $f(x)$.
It is clear that $x^*<a'$ and
\begin{equation}
x^*<a \quad {\rm if}\ 0\le p < \frac{1}{\nu} \frac{\varepsilon}{a'-a} , \qquad
x^*>a \quad {\rm if}\ p > \frac{1}{\nu} \frac{\varepsilon}{a'-a} .  
\end{equation}  
If $x^*<a$, the minimum of $f(x)$ in the interval $[a,a']$ is
at the point $x_{\min}=a'$.
If $x^*>a$, the minimum of $f(x)$ stays at $x_{\min}=a'$ until $f(a)=f(a')$,
i.e. until $p$ reaches the incompressibility pressure
\begin{equation}
p_i = \frac{\varepsilon}{a'-a} .
\end{equation}
Note that $p_i$ does not depend on $\nu$.
We conclude that if $0\le p<p_i$ the minimum of $f(x)$ is at $x_{\min}=a'$,
if $p=p_i$ the minimum of $f(x)$ is at the two boundary points $x_{\min}=a,a'$
and if $p>p_i$ the minimum of $f(x)$ is at $x_{\min}=a$.

\subsection{$0< p<p_i$} \label{Sec3.1}
First consider the case $0<p<p_i$, when the minimum of
$f(x)$ is at the point $x_{\min}=a'$.
By changing the reference point of integration from $a$ to $a'$ by
substituting the variables $x=a'-t$, the integral (\ref{Iaaps})
can be expressed as
\begin{equation} \label{Iaaps1}
I_{\beta}(a,a';p,s) = {\rm e}^{-a'(s+\beta p)} \int_0^{a'-a} {\rm d}t\,
{\rm e}^{(s+\beta p)t-\beta\varepsilon \left(\frac{t}{a'-a}\right)^{1/\nu}} . 
\end{equation}
Subsequent substitution
\begin{equation}
t=(a'-a)\left( \frac{u}{\beta\varepsilon} \right)^{\nu}
\end{equation}
implies
\begin{equation} \label{Iaaps2}
I_{\beta}(a,a';p,s) = {\rm e}^{-a'(s+\beta p)}
\frac{(a'-a)\nu}{(\beta\varepsilon)^{\nu}} \int_0^{\beta\varepsilon} {\rm d}u\,
u^{\nu-1} {\rm e}^{(s+\beta p)(a'-a)\left(\frac{u}{\beta\varepsilon}\right)^{\nu}-u} . 
\end{equation}
In the limit $\beta\to\infty$ the first term in the exponential under
integration is of the order $\beta^{1-\nu}$, so for $\nu>1$ the exponential
can be expanded in powers of this term.
In the leading order, we get
\begin{equation} \label{Iaaps3}
I_{\beta}(a,a';p,s) \mathop{\sim}_{\beta\to\infty} {\rm e}^{-a'(s+\beta p)}
\frac{(a'-a)\nu}{(\beta\varepsilon)^{\nu}} \Gamma(\nu) .
\end{equation}
This term is subleading in the limit $\beta\to\infty$ with respect to
the second term in Eq. (\ref{rovnica}) and we get
\begin{equation} \label{rovnica1}
\widehat{\Omega}(s+\beta p) \mathop{\sim}_{\beta\to\infty}
\frac{1}{\beta p} {\rm e}^{-a'(s+\beta p)} .
\end{equation}
As a result,
\begin{equation} \label{rovnica2}
\frac{\widehat{\Omega}(s+\beta p)}{\widehat{\Omega}(\beta p)}
\mathop{\sim}_{\beta\to\infty} {\rm e}^{-a's} .
\end{equation}

According to (\ref{lEoS}), the mean distance between NN particles
in the ground state is then expressed as
\begin{equation} 
l_0(p) = a' . 
\end{equation}
The correlation function at zero temperature
\begin{equation} \label{g0hr1}
g_0(x) = {\cal L}^{-1}\left\{a'\sum_{k=1}^\infty{\rm e}^{-a' k s}\right\}
= a' \sum_{k=1}^{\infty} \delta(x-k a') 
\end{equation}
corresponds to an equidistant ground state with spacing $a'$.
The isothermal susceptibility (\ref{chi}) vanishes at $T=0$,
\begin{equation}
\chi_0(p)=0 .
\end{equation}

\subsection{$p>p_i$} \label{Sec3.2}
In the case of $p>p_i$, the minimum of $f(x)$ is at the point $x_{\min}=a$.
Using the substitution of variables $x=a+t$, the integral (\ref{Iaaps})
can be expressed as
\begin{equation} \label{Iaaps4}
I_{\beta}(a,a';p,s) = {\rm e}^{-a(s+\beta p)} \int_0^{a'-a} {\rm d}t\,
{\rm e}^{-(s+\beta p)t-\beta\varepsilon \left(1-\frac{t}{a'-a}\right)^{1/\nu}} . 
\end{equation}
Subsequent substitution
\begin{equation}
t = \frac{u}{s+\beta p}
\end{equation}
implies
\begin{equation} \label{Iaaps5}
I_{\beta}(a,a';p,s) = \frac{{\rm e}^{-a(s+\beta p)}}{s+\beta p}
\int_0^{(a'-a)(\beta p+s)} {\rm d}u\,
{\rm e}^{-u-\beta\varepsilon\left( 1-\frac{u}{(a'-a)(\beta p+s)}\right)^{1/\nu}} . 
\end{equation}
In the limit $\beta\to\infty$, the second term in the exponential exponent
can be expanded in powers of $1/\beta$.
In the leading order, we thus get
\begin{equation} \label{Iaaps6}
I_{\beta}(a,a';p,s) \mathop{\sim}_{\beta\to\infty}
\frac{{\rm e}^{-a s -\beta (\varepsilon+p a)}}{\beta p}
\frac{1}{1-\frac{\varepsilon}{\nu(a'-a)p}} .
\end{equation}
Since $p a + \varepsilon < a' p$ for the considered pressures $p>p_i$,
this term is leading in the limit $\beta\to\infty$ with respect to
the second term in Eq. (\ref{rovnica}) and we obtain 
\begin{equation} \label{rovnica3}
\widehat{\Omega}(s+\beta p) \mathop{\sim}_{\beta\to\infty}
\frac{1}{\beta p\left[ 1-\frac{\varepsilon}{\nu(a'-a)p}\right]}
{\rm e}^{-a s-\beta(\varepsilon+pa)}.
\end{equation}
Consequently,
\begin{equation} \label{rovnica4}
\frac{\widehat{\Omega}(s+\beta p)}{\widehat{\Omega}(\beta p)}
\mathop{\sim}_{\beta\to\infty} {\rm e}^{-a s} .
\end{equation}

Given (\ref{lEoS}), the mean distance between NN particles
in the ground state is
\begin{equation} 
l_0(p) = a . 
\end{equation}
The correlation function at zero temperature
\begin{equation} \label{g0hr2}
g_0(x) = {\cal L}^{-1}\left\{a\sum_{k=1}^\infty{\rm e}^{-a k s}\right\}
= a \sum_{k=1}^{\infty} \delta(x-k a) 
\end{equation}
corresponds to an equidistant ground state with hard-core spacing $a$.
The isothermal susceptibility (\ref{chi}) vanishes at $T=0$,
\begin{equation}
\chi_0(p)=0 .
\end{equation}

\subsection{$p=p_i$} \label{Sec3.3}
As mentioned earlier, for $p=p_i$ the equality $f(a)=f(a')$ holds, that is
the function $f(x)$ has two equivalent minima at
the points $x_{\min}=a$ and $a'$.
Therefore, we divide the integral (\ref{Iaaps}) into two integrals,
one in the interval $x\in [a,b]$ and the other in the interval
$x\in [b,a']$, where $a<b<a'$.
The integral on the interval $x\in [a,b]$ can be treated analogously
to the integral in section \ref{Sec3.2}, which leads in the limit
$\beta\to\infty$ to the contribution (\ref{rovnica3})
with the substitution $p\to p_i$:
\begin{equation} \label{term1}
{\rm e}^{-\beta\varepsilon a'/(a'-a)} \frac{1}{\beta p_i} 
\frac{\nu}{\nu-1} {\rm e}^{-a s}.
\end{equation}
Note that this contribution does not depend on parameter $b$.
The integral on the interval $x\in [b,a']$ can be treated analogously
to the integral in section \ref{Sec3.1}, which leads in the limit
$\beta\to\infty$ to the contribution (\ref{Iaaps3}) with the
substitution $p\to p_i$.
This contribution is of order $\beta^{-\nu}$ and is therefore subleading
with respect to the one (\ref{term1}).
Adding the second term in the original relation (\ref{rovnica}),
finally we obtain
\begin{equation} \label{rovnicapi}
\widehat{\Omega}(s+\beta p_i) \mathop{\sim}_{\beta\to\infty}
{\rm e}^{-\beta\varepsilon a'/(a'-a)}\frac{1}{\beta p_i} 
\frac{1}{\nu-1} \left[ \nu {\rm e}^{-a s} + (\nu-1) {\rm e}^{-a's} \right] .
\end{equation}

The average distance of NN particles in the ground state is given by
\begin{equation} \label{l1/nu}
l_0(p_i) = \frac{\nu}{2\nu-1} a + \frac{\nu-1}{2\nu-1} a' .
\end{equation}
Note that the average distance is a kind of weighted average with probability
$\nu/(2\nu-1)$ of spacing $a$ and probability $(\nu-1)/(2\nu-1)$
of spacing $a'$.
Since the ratio
\begin{equation} \label{oo1/nu}
\lim_{\beta\to\infty}
\frac{\widehat{\Omega}(s+\beta p_i)}{\widehat{\Omega}(\beta p_i)}
= \frac{\nu}{2\nu-1} {\rm e}^{-a s} + \frac{\nu-1}{2\nu-1} {\rm e}^{-a' s} ,
\end{equation}
the ground-state correlation function is
\begin{equation} \label{g0xi1/nu}
g_0(x) = l_0(p_i) \sum_{j,k=0\atop (j,k)\ne (0,0)}^\infty
\left( \frac{\nu}{2\nu-1} \right)^j \left( \frac{\nu-1}{2\nu-1} \right)^k
{j+k \choose j} \delta(x - j a - k a') .
\end{equation}
Since there is no constant $c$ such that $g_0(x+c)=g_0(x)$,
$g_0(x)$ is not a periodic function and corresponds to
a disordered ground state.
This disordered ground state consists of a mixture of particle
configurations at discrete positions $j a + k a'$ with probability
$\left( \frac{\nu}{2\nu-1} \right)^j \left( \frac{\nu-1}{2\nu-1} \right)^k$
of $j$ spacings $a$ and $k$ spacings $a'$, the binomial coefficient
counts the number of their possible arrangements.

The isothermal susceptibility (\ref{chi}) at zero temperature is nonzero
and given by
\begin{equation} \label{chi01/nu}
\chi_0(p) =\nu(\nu-1)\left[\frac{a'-a}{\nu a +(\nu-1)a'}\right]^2 .
\end{equation}
This means that the disordered ground state is non-hyperuniform,
i.e., it allows significant large-scale density fluctuations as
in ordinary fluids at nonzero temperatures \cite{Torquato18}.
Interestingly, the disordered ground state,
which consists of a mixture of particle configurations at discrete
positions, resembles a disordered liquid rather than a crystal.

\begin{figure}
\centering
\begin{subfigure}{}
\includegraphics[clip,width=0.60\textwidth]{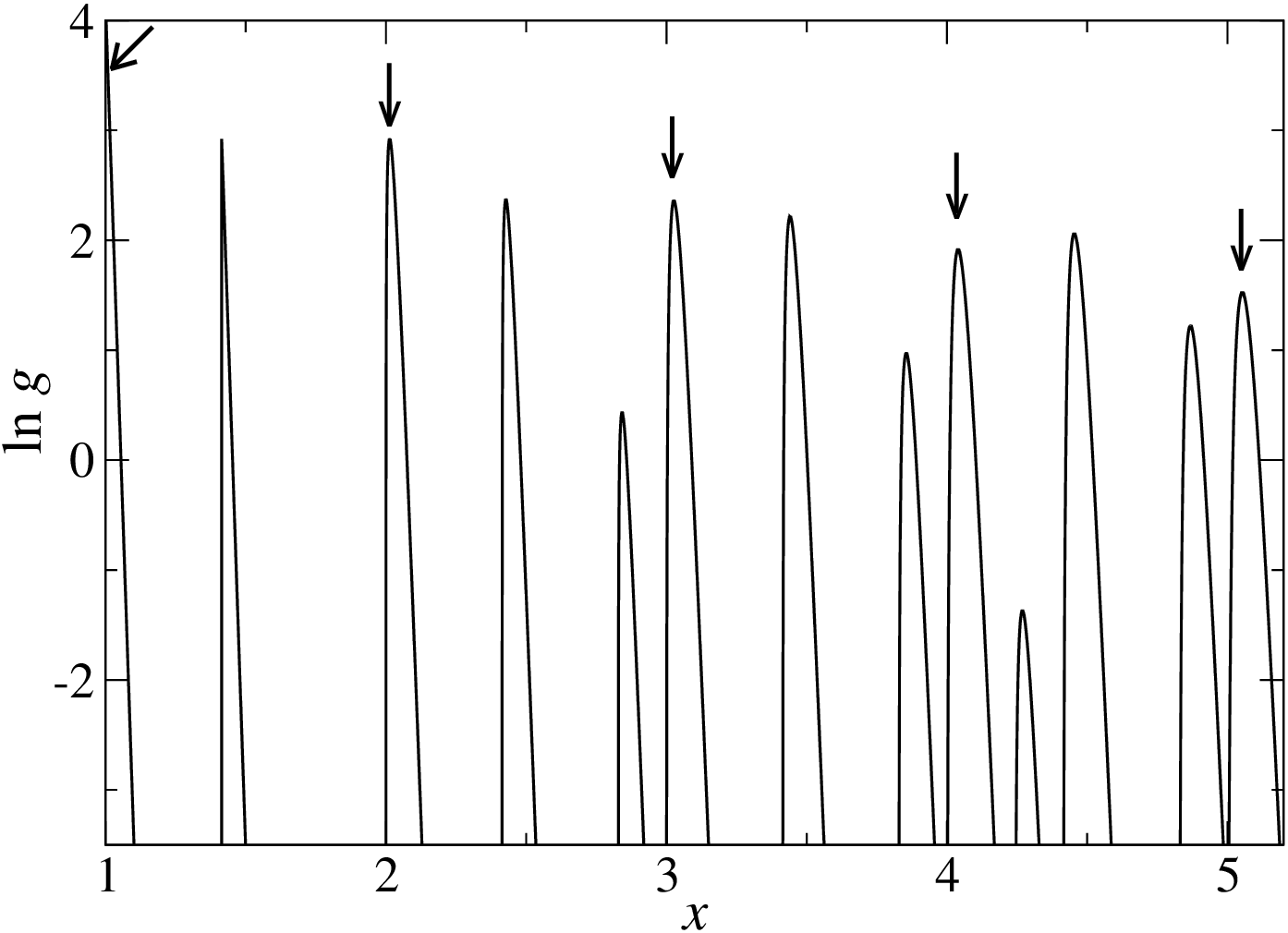}
\label{fig:subfig1A}
\end{subfigure}
\begin{subfigure}{}
\includegraphics[clip,width=0.60\textwidth]{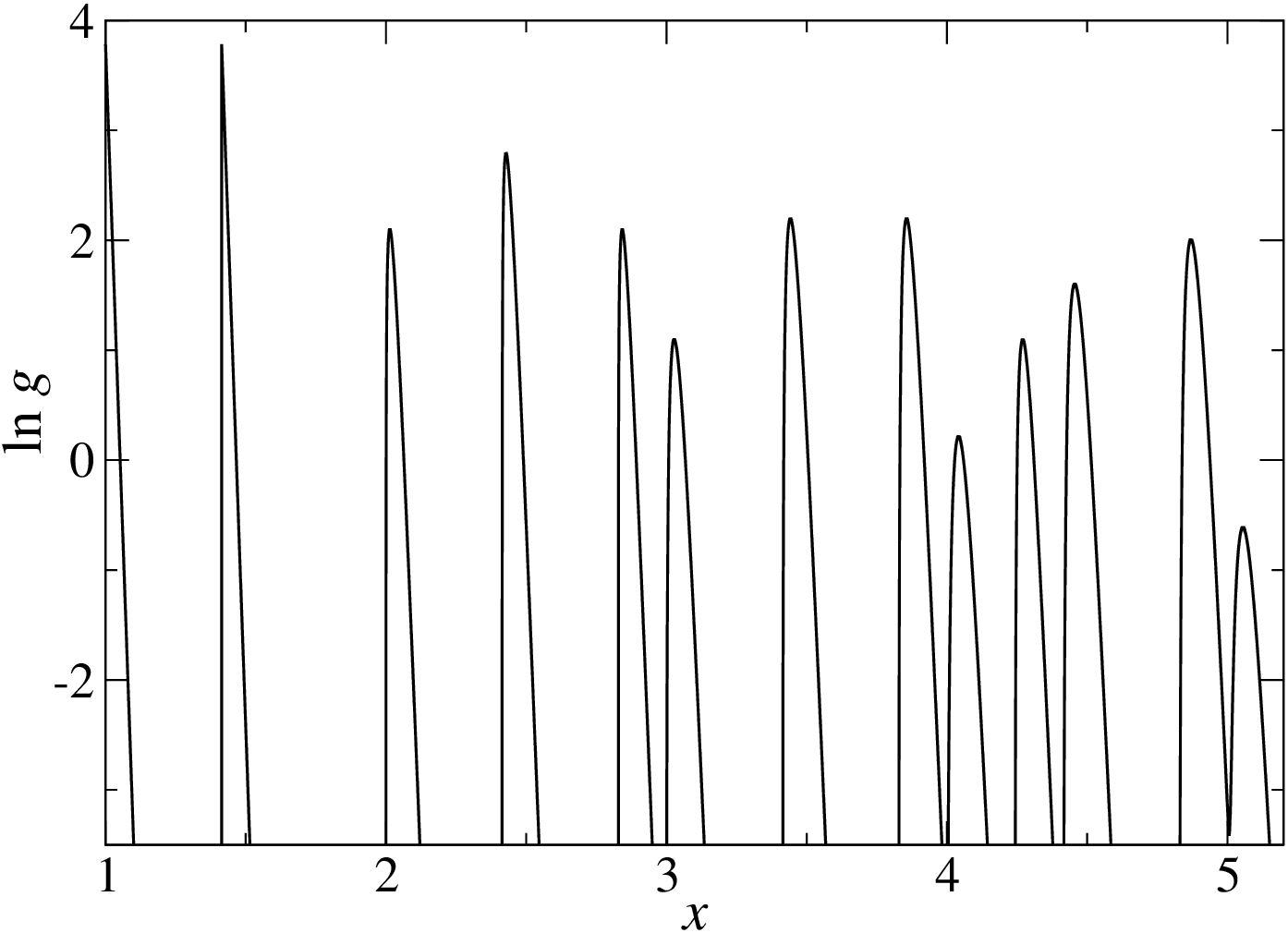}
\label{fig:subfig1B}
\end{subfigure}
\begin{subfigure}{}
\includegraphics[clip,width=0.60\textwidth]{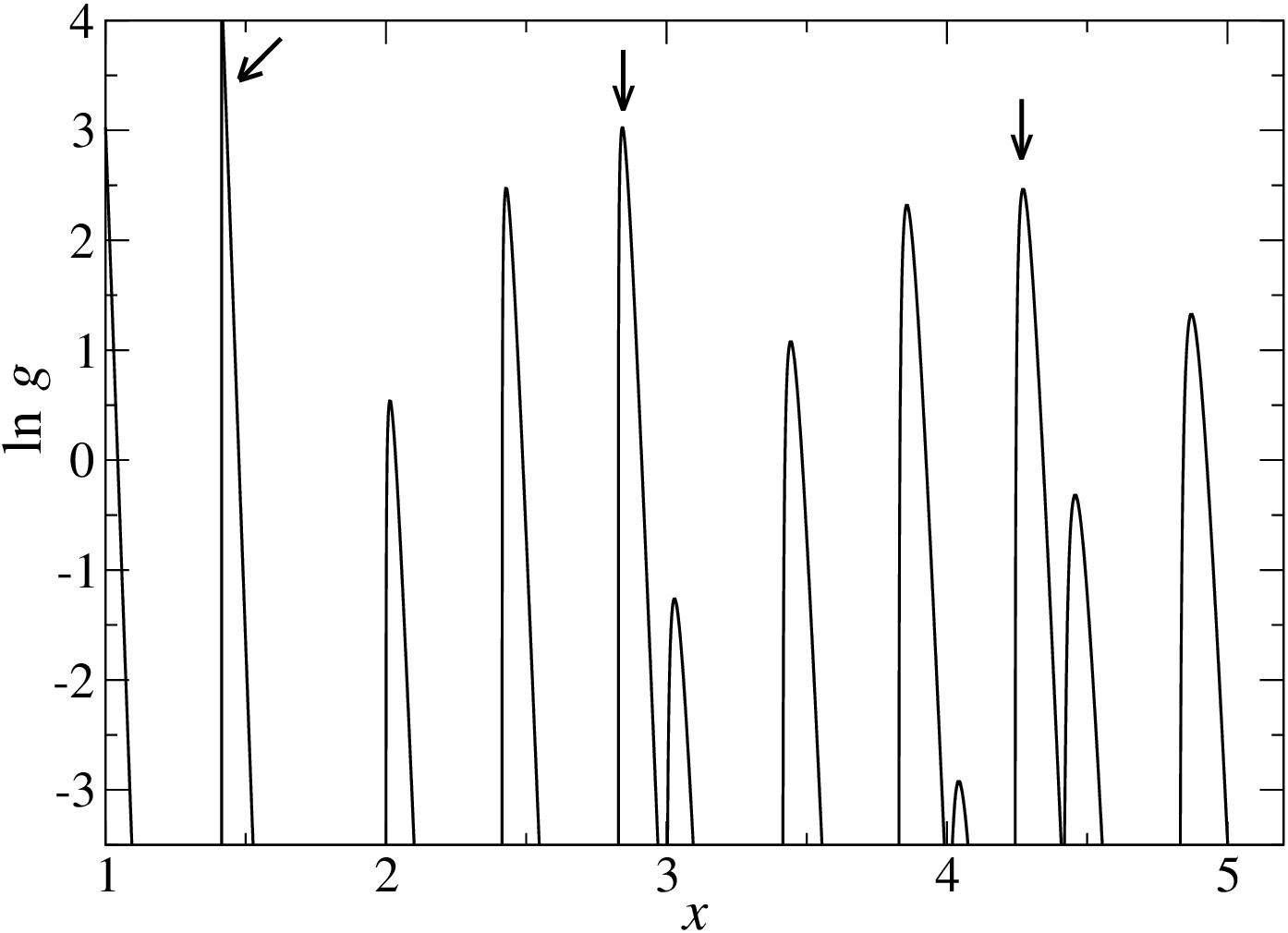}
\label{fig:subfig1C}
\end{subfigure}
\caption{Square shoulder potential $(\nu\to\infty)$ with
parameters $a=1$, $a'=\sqrt{2}$, $\varepsilon=1$ and incompressibility
pressure $p_i=\varepsilon/(a'-a)=1+\sqrt{2}$.  
Plots of correlation functions $g(x)$ in semi-logarithmic scale
at inverse temperature $\beta=30$ for three pressures: $p=p_i+0.1$
(top figure), $p=p_i$ (middle figure) and $p=p_i-0.1$ (bottom figure).
The arrows in the top and bottom figures indicate peaks originating in
equidistant ground states with periods $a=1$ and $a'=\sqrt{2}$,
respectively.
Arrows indicate peaks which survive in the ground-state limit
$\beta\to\infty$.}
\label{fig2}
\end{figure}

Since disordered ground states only occur at
one specific pressure value $p_i$, the fundamental question is
how robust they are against small, say thermal, perturbations. 
In particular, whether they have any effect on the correlation functions of
thermodynamic states at nonzero temperature for a wider range of pressures.
As a test model, we consider the square shoulder potential $(\nu\to\infty)$
with parameters $a=1$, $a'=\sqrt{2}$, $\varepsilon=1$,
the incompressibility pressure $p_i=\varepsilon/(a'-a)=1+\sqrt{2}$
and the ground-state correlation function
\begin{equation}
g_0(x) = \frac{a+a'}{2} \sum_{j,k=0\atop (j,k)\ne (0,0)}^{\infty}
\frac{1}{2^{j+k}} \left( {j+k\atop j} \right) \delta(x-ja-ka') .
\end{equation}  
In Fig. \ref{fig2} we show three plots of $g(x)$ in semi-logarithmic scale
at low temperature, namely the inverse temperature $\beta=30$.
In the middle figure, which corresponds to $p=p_i$, $g(x)$ exhibits
narrow peaks (instead of infinite delta function peaks) near the disordered
positions of the ground state 
$x=1,\sqrt{2},2,1+\sqrt{2},2\sqrt{2},3,2+\sqrt{2},1+2\sqrt{2},4,\ldots$.
In the top figure, corresponding to $p=p_i+0.1$, $g(x)$ exhibits
narrow peaks near equidistant (period $a$) ground-state positions
$x=1,2,3,4,\ldots$ indicated by arrows plus new peaks near
disordered ground-state positions
$x=\sqrt{2},1+\sqrt{2},2\sqrt{2},2+\sqrt{2},1+2\sqrt{2},\ldots$.
In the bottom figure, which corresponds to $p=p_i-0.1$, $g(x)$ exhibits
narrow peaks close to equidistant (period $a'$) ground-state positions
$x=\sqrt{2},2\sqrt{2},3\sqrt{2},\ldots$ marked by arrows plus
new peaks close to the disordered ground-state positions
$x=1,2,1+\sqrt{2},3,2+\sqrt{2},1+2\sqrt{2},4,\ldots$.
It is clear that the pair correlation function of the disordered ground
state extends its predictive power to thermodynamic states at nonzero
temperatures over a wider range of pressures around $p_i$.

\renewcommand{\theequation}{4.\arabic{equation}}
\setcounter{equation}{0}

\section{Linear ramp model with $\nu=1$} \label{Sec4}
The interaction potential of the linear ramp model is defined as
(\ref{general}) with 
\begin{equation} \label{lrm}
\varphi(x) = \varepsilon \frac{(a'-x)}{(a'-a)} , \qquad
a < x < a',
\end{equation}
where $\varepsilon>0$.
It is both convex and concave.
This core-softened potential describes water-like anomalies
\cite{Lomba07}.
Its formal 1D solution was discussed in \cite{Montero19}.

The Laplace transform (\ref{Om}) of the Boltzmann factor of
the linear potential (\ref{lrm}) is given by
\begin{equation} \label{EosRampl}
\widehat\Omega(s) = \frac{\beta\varepsilon\e^{-a' s}-(a'-a)s
\e^{-\beta\varepsilon-a s}}{s [\beta\varepsilon-(a'-a)s]} .
\end{equation}
The zero-temperature limit of the reciprocal density $l_0$ has exactly
the same form as the one for the square shoulder potential,
including the value of the impressibility pressure $p_i=\varepsilon/(a'-a)$.
Inserting the ratio
\begin{equation} \label{omom}
\lim_{\beta\to\infty}
\frac{\widehat{\Omega}(s+\beta p)}{\widehat{\Omega}(\beta p)}
= \left\{
\begin{array}{ll}
\displaystyle{{\rm e}^{-a' s}} & \mbox{if $\displaystyle{0<p<p_i}$,} \cr
\displaystyle{\frac{{\rm e}^{-a s}-{\rm e}^{-a's}}{(a'-a)s}} &
\mbox{if $\displaystyle{p=p_i}$,} \cr \displaystyle{{\rm e}^{-a s}} &
\mbox{if $\displaystyle{p>p_i}$,}  
\end{array} \right.   
\end{equation}
into the relation for the pair correlation function (\ref{G}), for $0<p<p_i$
and $p>p_i$ we obtain equidistant ground states with periods $a'$ and
$a$, respectively.
For $p=p_i$ we need to use the formula for the inverse Laplace transform
\begin{equation} \label{es/s}
{\cal L}^{-1}\left\{\frac{{\rm e}^{-\alpha s}}{s^k}\right\}
= \frac{(x-\alpha)^{k-1}}{(k-1)!}\theta(x-\alpha),\qquad k=1,2,\ldots ,
\end{equation}
where $\theta(x)$ is the Heaviside step function, to get
\begin{eqnarray} \label{g0lri}
g_0(x) & = & \frac{a+a'}{2}\sum_{j,k=0\atop (j,k)\ne (0,0)}^\infty
\frac{(-1)^k}{(a'-a)^{j+k}} {j+k \choose j} \nonumber \\ & &
\times\frac{[x-j a - k a']^{j+k-1}}{(j+k-1)!}
\theta\left( x -j a - k a'\right).
\end{eqnarray}
Note that the degree of the polynomials in $x$ attached to the step functions
surprisingly increases with distance $x$, but nevertheless $g_0(x)$
shows damped oscillations and, as we will see, tends to $1$ for
large $x$, as expected.

Respecting the non-strict inequality (\ref{condition}), the formula
for the pair correlation function (\ref{g0lri}) can be explicitly represented
for a series of finite intervals covering the whole distance
axis $x$.
Small values of distance $x$ can be easily determined as follows:
\begin{equation} \label{g0i}
g_0(x) = \left\{
\begin{array}{ll}
\displaystyle{\frac{a+a'}{2(a'-a)}} & \mbox{if $a\le x <a'$,} \cr
\displaystyle{0} & \mbox{if $\displaystyle{a'\le x<2a}$,} \cr
\displaystyle{\frac{a+a'}{(a'-a)^2}\left(\frac{x}{2}-a\right)} &
\mbox{if $\displaystyle{2a\le x<a+a'}$.} \cr \end{array} \right.   
\end{equation}
The next interval is
\begin{equation} \label{g0lriml}
g_0(x)= \frac{a'+a}{(a'-a)^2}\left(a'-\frac{x}{2}\right)\qquad
\mbox{if $a+a'\le x < \min\{ 2a',3a\}$.}
\end{equation}
Next, we must distinguish between the cases $2a'<3a$ and $2a'>3a$.
When $2a'<3a$ we have
\begin{equation}
g_0(x) = 0 \qquad \mbox{if $2 a' \le x < 3a$}
\end{equation}
and when $2a'>3a$ we have
\begin{eqnarray} \label{g0lrimq}
g_0(x) & = & \frac{a+a'}{4(a'-a)^3}
\left[ x^2-2x(2a+a')+9a^2-4aa'+4{a'}^2\right] 
\nonumber \\ & & \qquad \qquad \qquad \mbox{if $3a\le x< 2a'$.} 
\end{eqnarray}

\begin{figure}[t]
\begin{center}
\includegraphics[clip,width=0.84\textwidth]{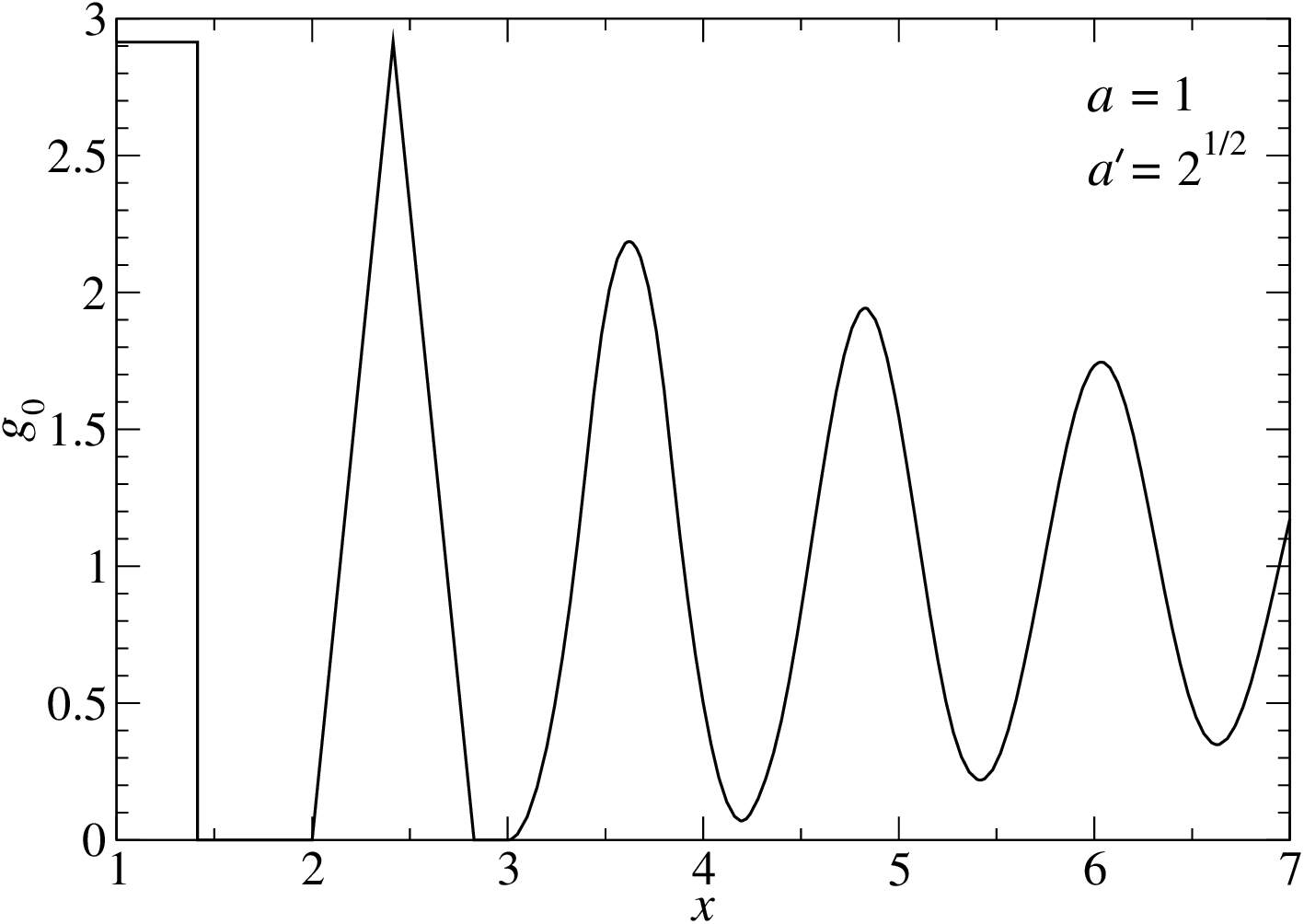}
\caption{Dependence of the ground-state correlation function $g_0$
on the distance $x$ for the linear ramp model (\ref{lrm}) with parameters
$a=1$, $a'=\sqrt{2}$, $\varepsilon=1$ and under the incompressibility
pressure $p_i=1+\sqrt{2}$.
For this model, the inequality $2a'<3a$ is satisfied.
See the text for a detailed description of the plot.}
\label{fig3}
\end{center}
\end{figure}

The plot of $g_0$ versus distance $x$ for the linear ramp model (\ref{lrm})
with parameters $a=1$ and $a'=\sqrt{2}$, such that $2a'<3a$,
is shown in Fig. \ref{fig3}.
There are two intervals of distance when $g_0$ is zero,
namely $\sqrt{2}<x<2$ and $2\sqrt{2}<x<3$.
The incommensurate choice of irrational $a'$ together with rational $a=1$
ensures that the values $j a+k a'$ in the argument of the Heaviside
functions in (\ref{g0lri}) are unique for any pair of non-negative
integers $j$ and $k$.
It is seen that damped oscillations with a period of approximately
$(a+a')/2$ tend toward $g_0=1$ at large $x$.

\begin{figure}[t]
\begin{center}
\includegraphics[clip,width=0.84\textwidth]{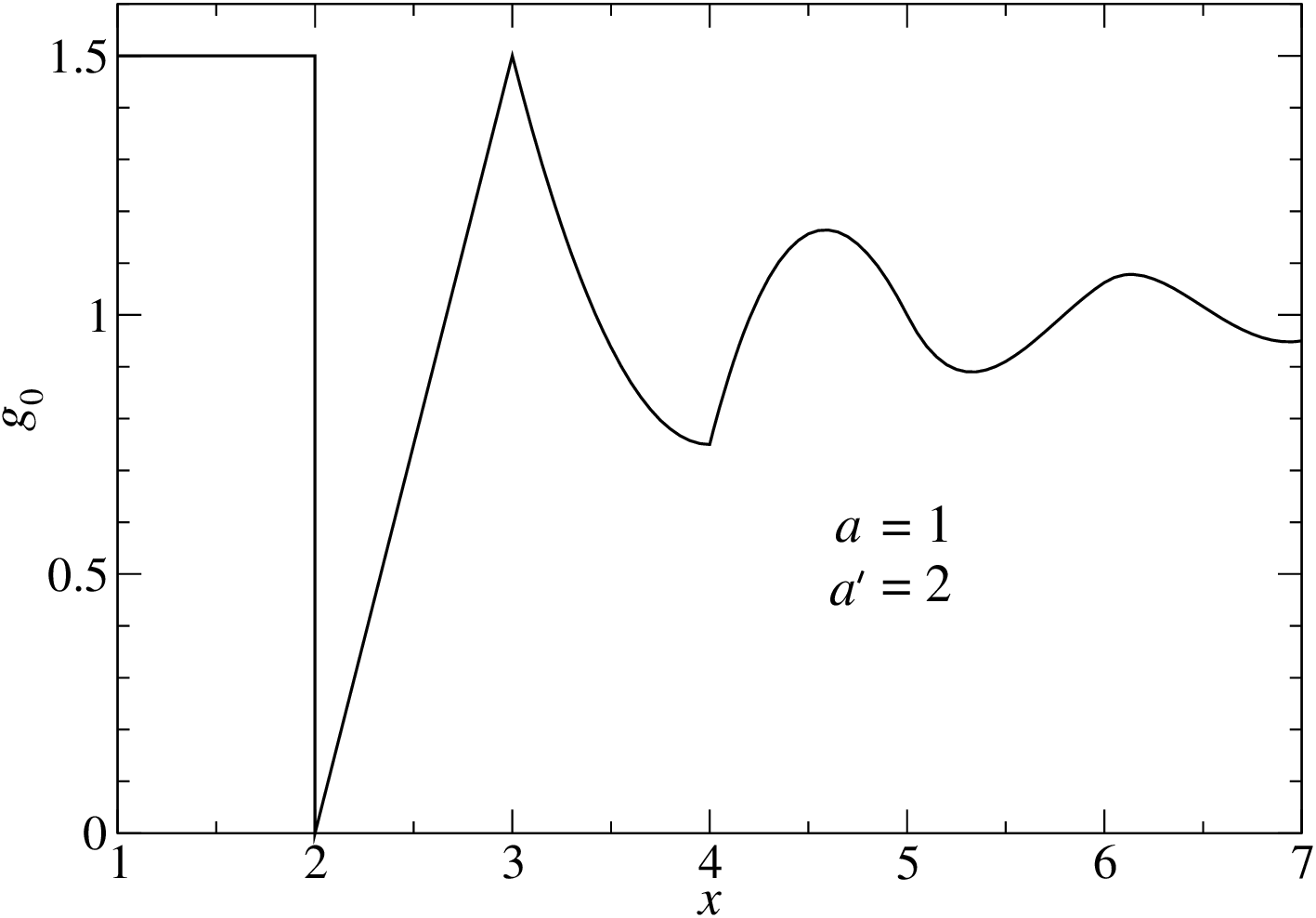}
\caption{Dependence of the ground-state correlation function $g_0$
on the distance $x$ for the linear ramp model (\ref{lrm}) with parameters
$a=1$, $a'=2$, $\varepsilon=1$ and under the incompressibility pressure $p_i=1$.
The inequality $2a'>3a$ holds in this model.
A detailed description of the plot can be found in the text.}
\label{fig4}
\end{center}
\end{figure}

The plot of $g_0$ versus distance $x$ for the linear ramp model (\ref{lrm})
with parameters $a=1$ and $a'=2$, such that $2a'>3a$,
is shown in Fig. \ref{fig4}.
The interval of vanishing $g_0$ contracts to the point $x=2a=a'=2$.
For larger distances $x$, $g_0(x)$ exhibits damped oscillations with
a period of approximately $(a+a')/2$ and tends to $g_0=1$.

The value of the isothermal susceptibility at zero temperature is given by
\begin{equation} \label{chi0lr}
\chi_0 = \left\{ \begin{array}{ll}
\displaystyle{0} & \mbox{if $\displaystyle{p\ne p_i}$,} \cr
\displaystyle{\frac{1}{3}\left(\frac{a'-a}{a'+ a}\right)^2} &
\mbox{if $\displaystyle{p=p_i}$.}
\end{array} \right.
\end{equation}
This means that, as in the case of strictly concave interaction potentials,
disordered ground states are non-hyperuniform \cite{Torquato18}.
Note that the formula (\ref{chi0lr}) for $\nu=1$ is not the limiting
$\nu\to 1$ case of the previous formula (\ref{chi01/nu}) valid for $\nu>1$.  

\begin{figure}
\centering
\begin{subfigure}{}
\includegraphics[clip,width=0.60\textwidth]{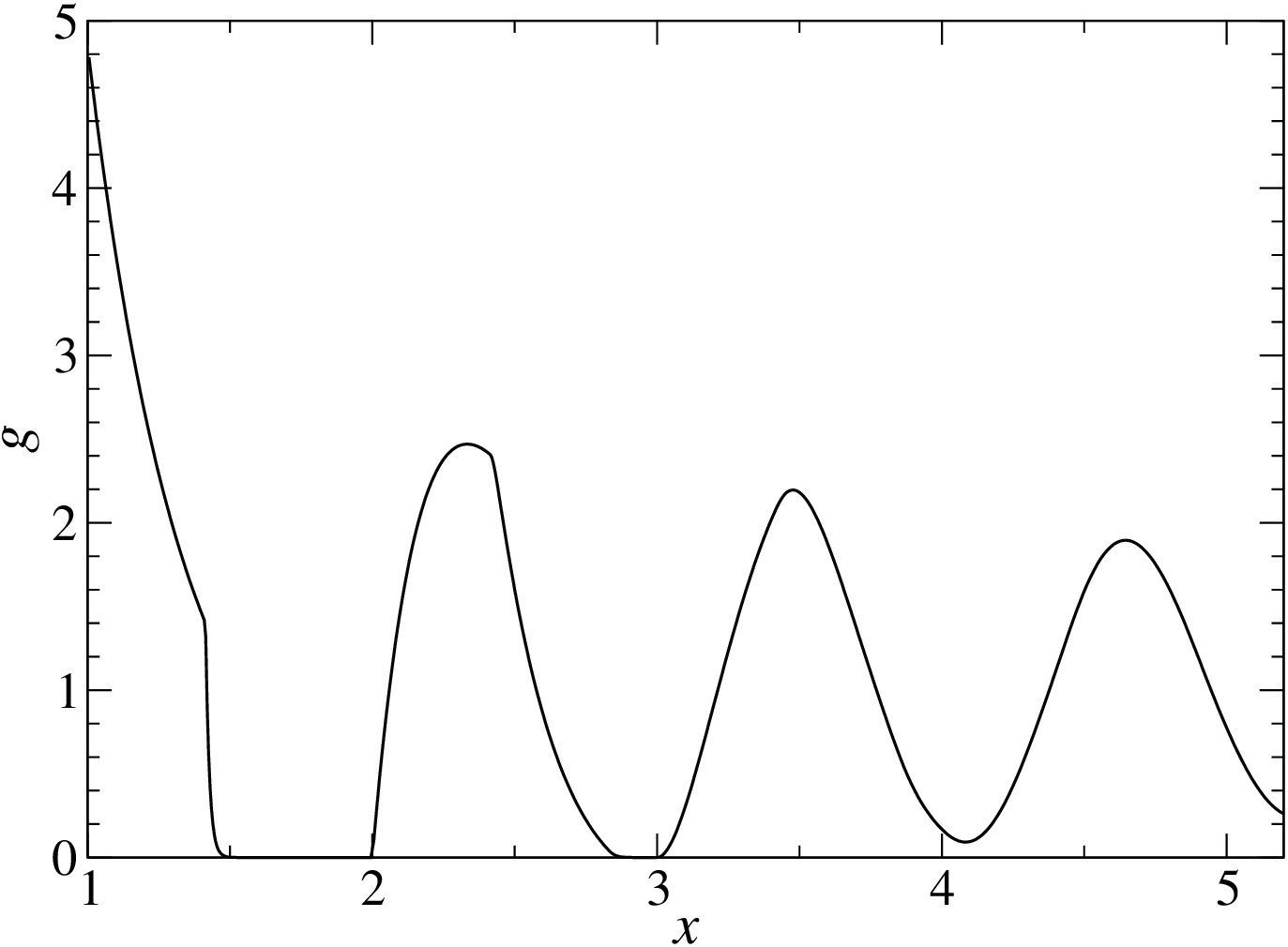}
\label{fig:subfig2A}
\end{subfigure}
\begin{subfigure}{}
\includegraphics[clip,width=0.60\textwidth]{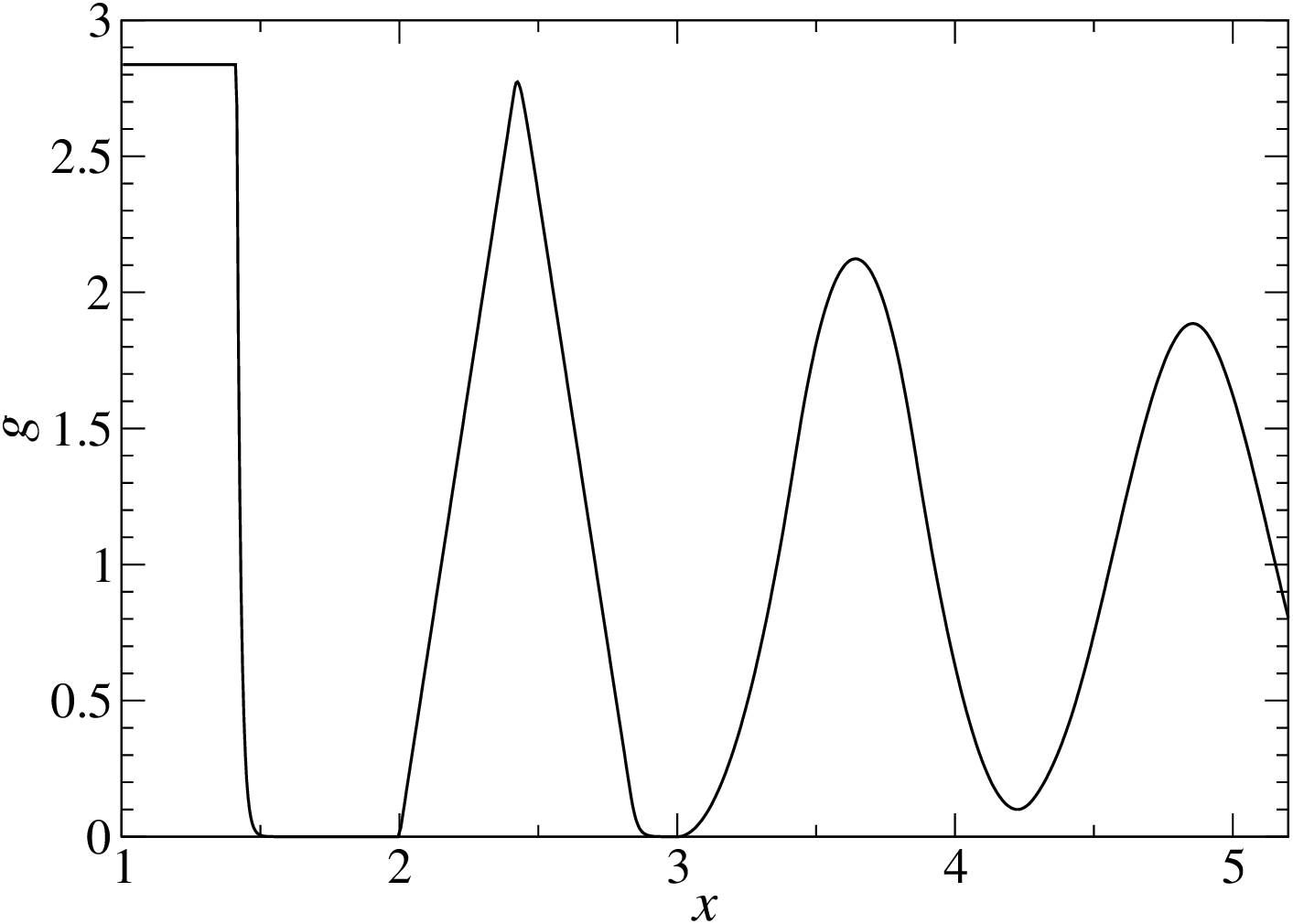}
\label{fig:subfig2B}
\end{subfigure}
\begin{subfigure}{}
\includegraphics[clip,width=0.60\textwidth]{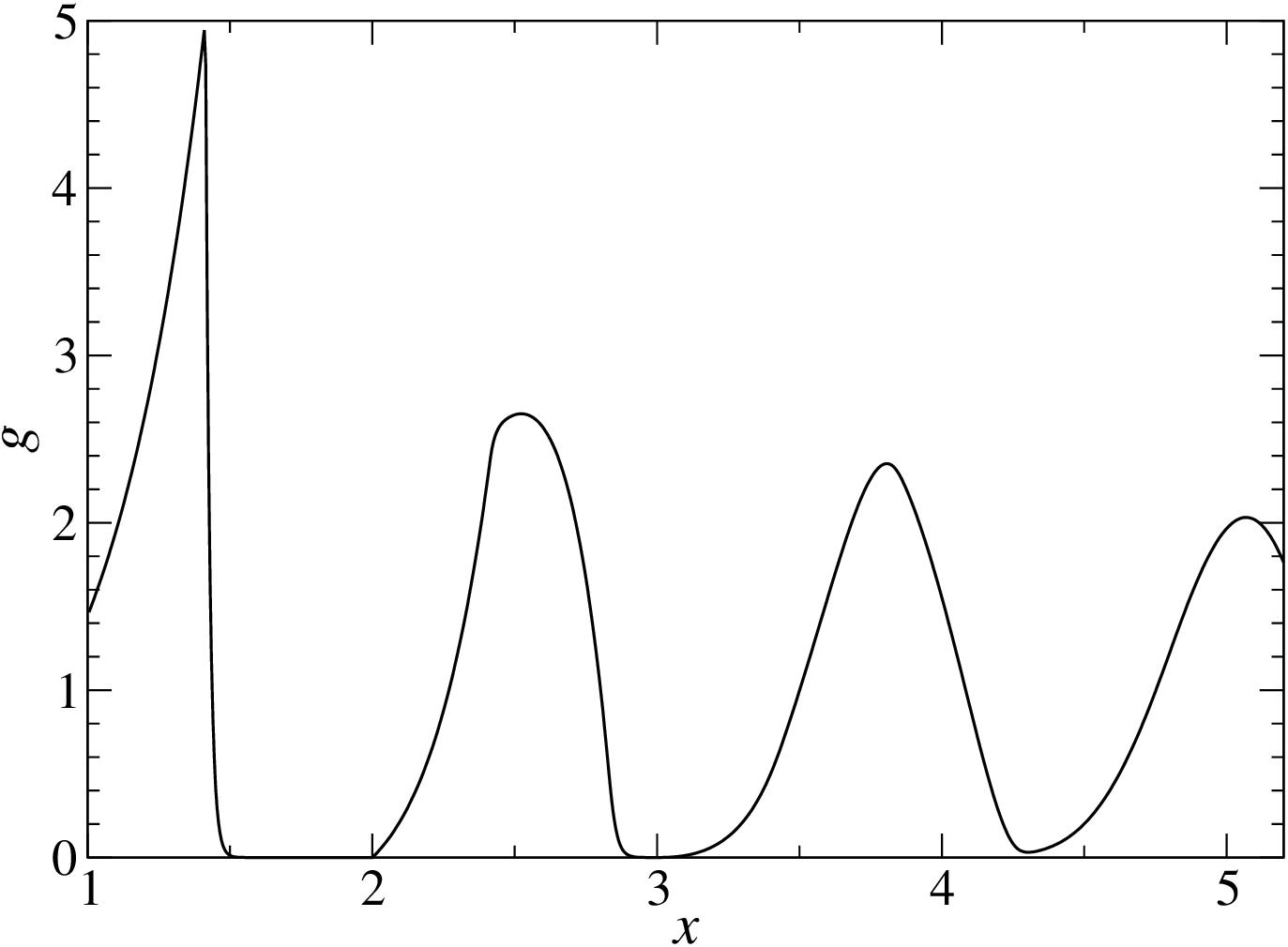}
\label{fig:subfig2C}
\end{subfigure}
\caption{The linear ramp potential with parameters
$a=1$, $a'=\sqrt{2}$, $\varepsilon=1$ and the incompressibility pressure
$p_i=\varepsilon/(a'-a)=1+\sqrt{2}$.  
Plots of the correlation functions $g(x)$ in semi-logarithmic scale
at inverse temperature $\beta=30$ for three pressures: $p=p_i+0.1$
(top figure), $p=p_i$ (middle figure) and $p=p_i-0.1$ (bottom figure).}
\label{fig5}
\end{figure}

To test the disordered ground state against small thermal
perturbations, we consider the linear ramp potential with parameters $a=1$,
$a'=\sqrt{2}$, $\varepsilon=1$ and the incompressibility pressure
$p_i=\varepsilon/(a'-a)=1+\sqrt{2}$.
In Fig. \ref{fig5} we show three plots of $g(x)$ in semi-logarithmic scale
at inverse temperature $\beta=30$ (low temperature).
In the middle figure, which corresponds to $p=p_i$, $g(x)$ exhibits a behavior
similar to that in the ground state shown in Fig. \ref{fig3}.
In the top figure, which corresponds to $p=p_i+0.1$, $g(x)$ exhibits
a slight deformation compared to the plot of $g(x)$ in the middle figure.
This $g(x)$ is very different from the equidistant ground state with
spacing $a=1$ that occurs at pressure $p=p_i+0.1$.
In the bottom figure, which corresponds to $p=p_i-0.1$, $g(x)$ also shows
only a slight change compared to the $g(x)$ plot in the middle figure.
As before, it is very different from the equidistant ground state with
spacing $a'=\sqrt{2}$ that occurs at pressure $p=p_i-0.1$.
We conclude that the predictive power of the pair correlation function
of the disordered ground state persists even at nonzero temperatures
and covers a wider range of pressures around $p_i$.

\renewcommand{\theequation}{5.\arabic{equation}}
\setcounter{equation}{0}

\section{Strictly convex interaction potentials with $0<\nu<1$} \label{Sec5}
The interaction potential
\begin{equation} \label{finu}
\varphi(x) = \varepsilon \left[\frac{(a'-x)}{(a'-a)}\right]^{1/\nu} , \qquad
a<x< a',
\end{equation}
with $0<\nu<1$ is strictly convex.
The derivative $\partial\varphi(x)/\partial x$ is continuous for
$a<x\le a'$ and vanishes for both $x\to {a'}^-$ and $x\to {a'}^+$.
The EoS for the equidistant ground state with spacing $l_0(p)$ can be
thus derived by minimizing the function $f(x)$ (\ref{fx}), namely
from the formula
\begin{equation} \label{derener}
p = - \frac{\partial\varphi(l_0)}{\partial l_0} ,
\end{equation}
see Ref. \cite{Travenec25}.
Explicitly,
\begin{equation} \label{l0p}
l_0(p) = a' - \left( \frac{p}{p_i} \right)^{\frac{\nu}{1-\nu}} (a'-a) ,
\end{equation}
where
\begin{equation} \label{pi}
p_i = \frac{\varepsilon}{\nu (a'-a)} .
\end{equation}
In the limit $p\to 0^+$ we have $l_0=a'$.
As $p$ increases, $l_0(p)$ continuously decreases from $a'$ to the hard-core
value $a$, which is reached at $p=p_i$.
For $p\ge p_i$ $l_0(p)=a$, which means that $p_i$ given by (\ref{pi}) is
the incompressibility pressure which, unlike concave potentials,
now depends on $\nu$.
Since all ground states are equidistant with unique spacing, no disordered
ground states were observed.

The correlation function at zero temperature is given by
\begin{equation} \label{g0q}
g_0(x) = \left\{
\begin{array}{ll}
l_0(p) \sum_{j=1}^{\infty} \delta\left[ x-j l_0(p)\right] 
& \mbox{if $\displaystyle{0<p<p_i}$,} \cr
\displaystyle{a\sum_{j=1}^\infty \delta(x-j a)} &
\mbox{if $\displaystyle{p\ge p_i}$.} 
\end{array} \right.   
\end{equation}
The susceptibility at zero temperature is zero for all $0<\nu<1$
and pressures $p>0$.

\renewcommand{\theequation}{6.\arabic{equation}}
\setcounter{equation}{0}

\section{Conclusion} \label{Sec6}
1D fluids of particles with hard cores of diameter $a$ and soft
NN interaction potentials (\ref{nu}) with finite range $a'$ such that
$a'\le 2a$ are exactly solvable at any temperature $T$ and pressure $p>0$.
This allowed us to determine the ground state of particle systems as
the $T\to 0$ limit of exact expressions for EoS and correlation function.
Most systems exhibit trivially equidistant ground states.
The only exceptions are concave interaction potentials with $\nu\ge 1$, which
exhibit disordered ground states at a special value of the
incompressibility pressure $p_i=\varepsilon/(a'-a)$.
Strictly concave potentials with $\nu>1$ are studied in section \ref{Sec3}.
Analysis of the limit $\beta\to\infty$ of the exact equations using
the steepest descent method shows that the equidistant ground state
with spacing $a'$ for $0<p<p_i$ jumps to the one with spacing
$a$ for $p>p_i$.
The coexistence of microscopic configurations of NN particles with different
interparticle distances $a$ and $a'$ is possible at incompressibility
pressure $p_i$.
This is the key reason for the existence of disordered ground states, which
appear as weighted mixtures of pairs of NN particles with spacing $a$ or
$a'$, see formulas for EoS (\ref{l1/nu}) and pair correlation function
(\ref{g0xi1/nu}).
The most interesting results were obtained for the linear ramp potential
with $\nu=1$, which is both concave and convex, see section \ref{Sec4}.
Up to certain gaps, continuous distances between particles are allowed
for this potential. 
Although the mean distance between NN particles $l_0(p_i)=(a+a')/2$ is
simple, the correlation function (\ref{g0lri}) is given by an infinite
sum of Heaviside step functions multiplied by polynomials in the distance,
whose degree increases with distance.
Nevertheless, the correlation function exhibits damped oscillations and
a slope towards $1$ for large distances, as it should be,
see Figs. \ref{fig3} and \ref{fig4}.
The isothermal susceptibility $\chi_0(p_i)$ is nonzero for all
disordered ground states.
Therefore, they are non-hyperuniform and resemble disordered liquids 
at nonzero temperatures.
Strictly convex potentials with $0<\nu<1$, studied in
section \ref{Sec5}, exhibit a continuous change in equidistant
spacing with pressure $p$, from $a'$ at $p\to 0$ to $a$ at
the incompressibility pressure $p_i=\varepsilon/[\nu(a'-a)]$.
Since there is no pressure at which two equidistant ground states with
different spacings coexist, there are no disordered ground states
for strictly convex potentials.

\ack
This work was supported by the Slovak Research and Development Agency under
the Contract no. APVV-24-0091 and VEGA Grant No. 2/0089/24.

\end{document}